\documentclass[a4paper]{jpconf}
\usepackage{graphicx,amsmath,amssymb}
\newcommand{\secdec}{{\textsc{SecDec}}}
\newcommand{\gosam}{{\textsc{GoSam}}}
\newcommand{\python}{{\texttt{python}}}
\def\eps{\epsilon}
\newcommand{\nn}{{\nonumber}}
\newcommand{\bea}{\begin{eqnarray}}
\newcommand{\eea}{\end{eqnarray}}

\usepackage{cite}

\pdfoutput=1

\begin{document}
\title{Numerical multi-loop calculations: tools and applications}

\author{S.~Borowka$^1$, G.~Heinrich$^{2,}$\footnote[3]{Speaker; presented at the conference ACAT 2016, Valparaiso, Chile, January 2016.}, S.~Jahn$^2$, S.~P.~Jones$^2$, M.~Kerner$^2$, J.~Schlenk$^2$, T.~Zirke$^2$}

\address{$^1$ Institute for Physics, University of Z{\"u}rich, Winterthurerstr.190, 8057 Z{\"u}rich,
Switzerland}
\address{$^2$ Max Planck Institute for Physics, F\"ohringer Ring 6, 80805 M{\"u}nchen, Germany}

\ead{sborowka@physik.uzh.ch, gudrun@mpp.mpg.de, sjahn@mpp.mpg.de, sjones@mpp.mpg.de, kerner@mpp.mpg.de, jschlenk@mpp.mpg.de, zirke@mpp.mpg.de}

\begin{abstract}
We report on the development of tools to calculate loop integrals and amplitudes 
beyond one loop. 
In particular, we review new features of the program \secdec{} which can be used for the 
numerical evaluation of parametric integrals like multi-loop integrals. 
\end{abstract}

\section{Introduction}

Precision calculations for the LHC and future colliders are entering more and more the domain of 
predictions beyond the next-to-leading order (NLO). 
The rapid progress was mainly triggered by the development of better calculational methods and tools,
but also by the development of more powerful computing resources.
Both aspects lay the groundwork allowing to move towards the automation of calculations beyond NLO.

In these proceedings we will focus on the automation of two-loop calculations, in particular two-loop integrals
with several mass scales. While the latter are very difficult to calculate
analytically, numerical methods may offer a solution.
However, the numerical integration can only succeed if the various types of singularities which may occur 
in a two-loop amplitude can be isolated or evaded beforehand, preferably in an automated way. 
The program \secdec~\cite{Carter:2010hi,Borowka:2012yc,Borowka:2013cma,Borowka:2015mxa} is able to 
perform this task, and to subsequently integrate the resulting finite parameter integrals numerically. 
In the following we will describe the program \secdec{} and some of its new features,
as well as its embedding into a framework to generate two-loop amplitudes.

\section{Amplitude generation at one\,- and two loops}

\begin{figure}
\begin{center}
\includegraphics[width=10cm]{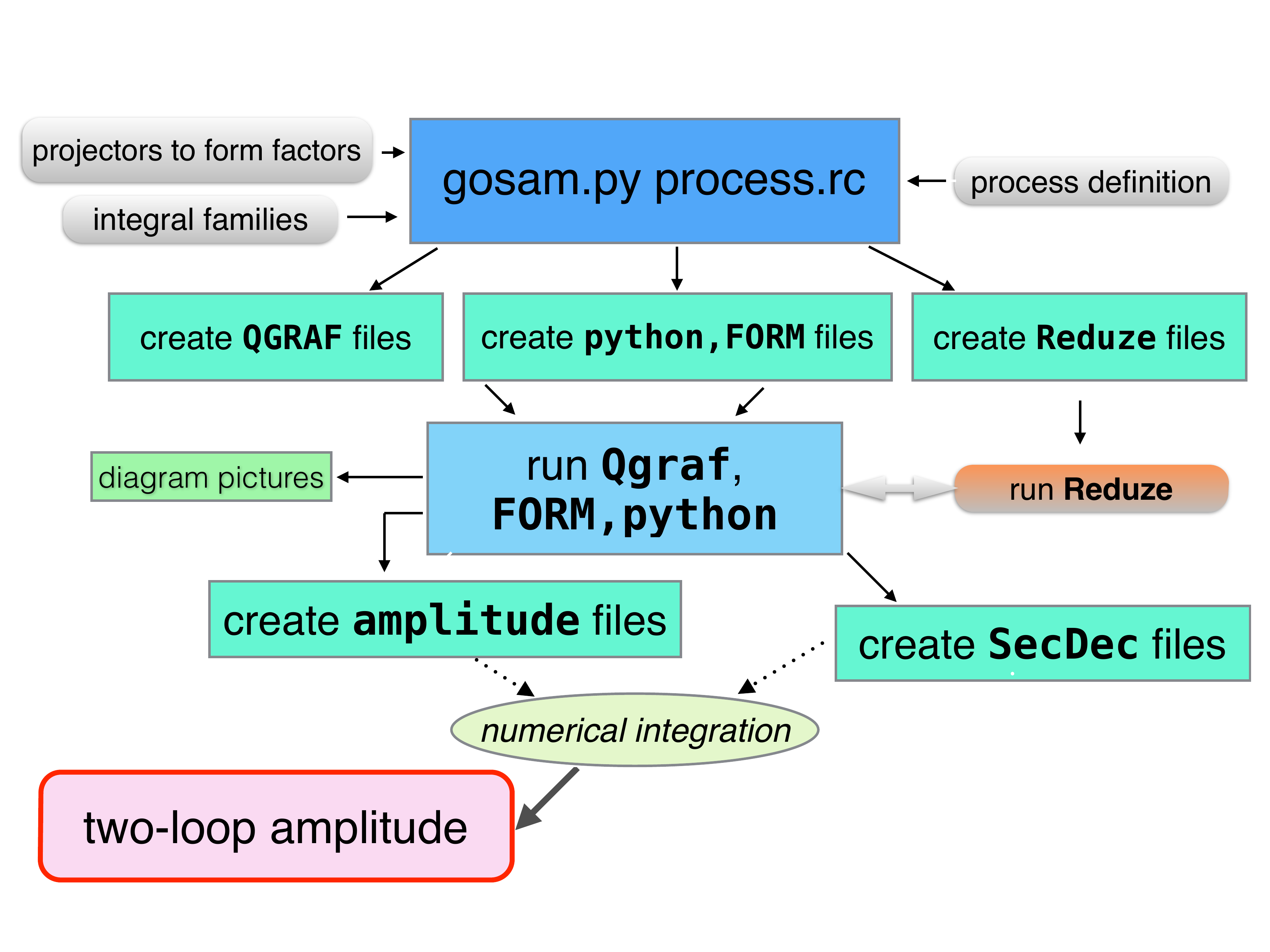}
\end{center}
\caption{Building blocks of \gosam-2loop. Grey boxes denote user
  input. The reduction to master integrals is run separately.
	\label{fig:gosam2} }
\end{figure}

The steps to be performed in order to generate a virtual amplitude (at in principle any loop order) 
in an approach based on Feynman diagrams are roughly the following: 
\begin{enumerate}
\item generation of the contributing diagrams in terms of algebraic expressions, 
based on the Feynman rules derived from a Lagrangian for a certain model,
\item casting the expression into a form ``coefficients $\otimes$ (tensor) loop integrals'',
\item reduction of the loop integrals to a set of master integrals (ideally, this set is a minimal basis 
in loop integral space) times coefficients, where the latter coefficients will depend on the 
dimension $D$ of loop momentum space,
\item evaluation of the master integrals, combination with the coefficients forming the amplitude.
\end{enumerate}
We are constructing a program package which is designed to perform the steps listed above in an automated way.
It partly builds on the program package \gosam~\cite{Cullen:2011ac,Cullen:2014yla} for automated one-loop 
calculations; for example it also uses {\sc Qgraf}~\cite{Nogueira:1991ex} for the diagram generation. 
However, the reduction step and the evaluation of the master integrals are much more challenging than at one loop,
as the integral basis is usually not known beforehand, and the master integrals may not be available in analytic form. 
Within the  \gosam-2loop framework, we constructed an interface to the 
reduction program {\sc Reduze}~\cite{Studerus:2009ye,vonManteuffel:2012np}. 
The reduction to master integrals is then run separately, and the resulting database is imported back into  
\gosam-2loop. The latter also contains an interface to the  program \secdec{}
which is used to evaluate the integrals numerically. 
The flowchart for \gosam-2loop, shown in Fig.~\ref{fig:gosam2}, illustrates the main steps.

\section{The program \secdec}

The program
\secdec\,\cite{Carter:2010hi,Borowka:2012yc,Borowka:2013cma,Borowka:2015mxa}
is based on the sector decomposition algorithm described in
\cite{Binoth:2000ps,Heinrich:2008si}. 
It can be applied to multi-loop integrals as well as more general
parametric integrals which contain dimensionally regulated end-point singularities.
Other public implementations of sector decomposition are described 
 in~\cite{Bogner:2007cr,Gluza:2010rn,Smirnov:2008py,Smirnov:2013eza,Smirnov:2015mct}. 

\subsection{Feynman Integrals}

We consider scalar multi-loop integrals in order to simplify the
notation. The case of tensor integrals is analogous, for details we
refer to~\cite{Heinrich:2008si}.
A scalar Feynman integral $I$ in $D$ dimensions 
at $L$ loops with  $N$ propagators, where 
the propagators can have arbitrary, not necessarily integer powers $\nu_j$,  
has the following representation in momentum space:
\begin{eqnarray}\label{eq0}
&&I=\frac{\mu^{4-D}}{i\pi^{\frac{D}{2}}} \int d^D k_l\;\frac{1}{\prod\limits_{j=1}^{N} P_{j}^{\nu_j}(\{k\},\{p\},m_j^2)}\;,\nn\\
&&P_j(\{k\},\{p\},m_j^2)=q_j^2-m_j^2+i\delta\;,
\end{eqnarray}
where the $q_j$ are linear combinations of external momenta $p_i$ and loop momenta $k_l$.
Introducing Feynman parameters and carrying out the momentum
integrations leads to
\begin{eqnarray} 
G&=&\frac{(-1)^{N_{\nu}}}{\prod_{j=1}^{N}\Gamma(\nu_j)}\Gamma(N_{\nu}-LD/2)\int
\limits_{0}^{\infty} 
\,\prod\limits_{j=1}^{N}dx_j\,\,x_j^{\nu_j-1}\,\delta(1-\sum_{l=1}^N x_l)
\frac{{\cal U}(\vec x)^{N_{\nu}-(L+1) D/2}}
{{\cal F}(\vec x,\{s_{ij}, m_i^2\})^{N_\nu-L D/2}}\;,\label{eq:FU} 
\end{eqnarray}
with $N_\nu=\sum_{j=1}^N\nu_j$. The graph polynomial ${\cal U}$ is a positive-semidefinite function of the Feynman parameters only, 
while ${\cal F}$ is a function of Feynman parameters and the kinematic invariants characterising the integral. 
In the region where all Lorentz invariants formed from external momenta are negative, 
which we call the {\em Euclidean region}, 
${\cal F}$ is also a positive semi-definite function 
of the Feynman parameters $x_j$ and the invariants.  
If some of the invariants (masses, squared external momenta) are zero,
the vanishing of  ${\cal F}$  
may lead to an infrared (soft or collinear) divergence.
Thus it depends on the {\em kinematics} and not only on the topology (as in the ultraviolet case) 
whether a zero of ${\cal F}$ leads to a divergence.
If some of the kinematic invariants are  positive, ${\cal F}$ can also vanish at points in Feynman parameter space which 
are not the endpoints of the parametric integration. To be able to deal with such thresholds (or pseudo-thresholds), 
a deformation of the integration contour into the complex plane has been implemented~\cite{Borowka:2012yc}.

The program \secdec{} can construct the graph polynomials ${\cal U}$ and ${\cal F}$ automatically.
Acting on polynomials raised to some power which usually contains the dimensional regulator $\eps$, 
it will  extract the poles in $1/\eps$ and evaluate their coefficients numerically, as explained in the next subsection.

\subsection{Basic structure of the program}

\begin{figure}
\begin{center}
\includegraphics[width=0.7\textwidth]{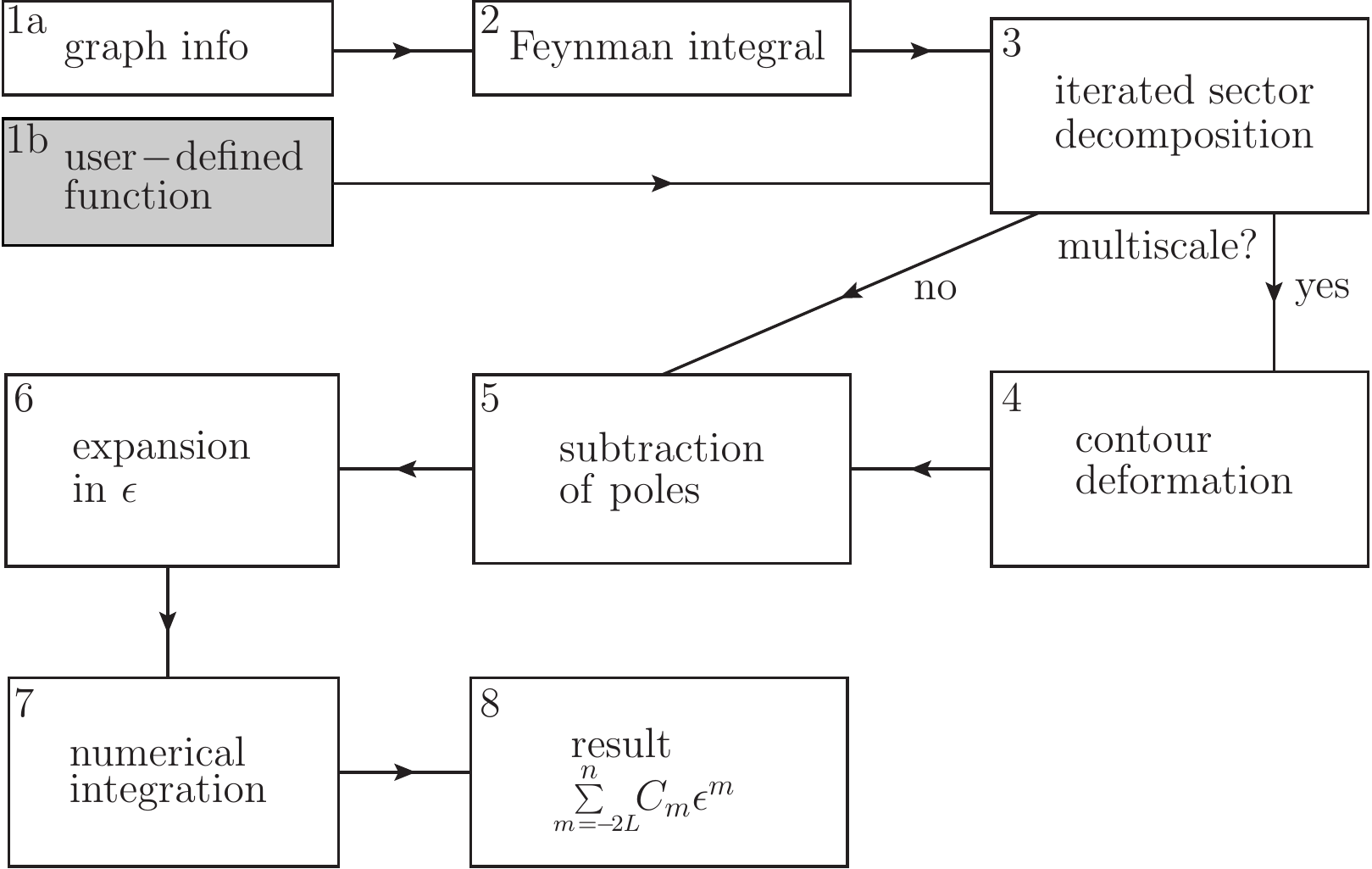}
\end{center}
\caption{Flowchart showing the main steps the program performs to produce the 
	numerical result as a Laurent series in $\epsilon$.
	\label{fig:structure} }
\end{figure}

The program consists of two main parts, one being designed for loop integrals, 
to be found in a directory called  {\tt loop}, the other one for more general parametric integrals, 
located in a directory called {\tt general}. 
The procedure to isolate the poles in the regulator $\eps$ and to do the subtractions and integrations 
is very similar in the two branches. However, only the {\tt loop} part contains the 
possibility of contour deformation, because only for loop integrals the analytic continuation 
can be performed in an automated way, following Feynman's ``$i \delta$" prescription.
The basic flowchart of the program is shown in Fig.~\ref{fig:structure}.

\subsection{New features}

\secdec-3.0~\cite{Borowka:2015mxa} has a number of new features, aiming more towards
applications in which the program is used as part of the calculation of a
full virtual amplitude, rather than serving only as a numerical check
for master integrals which are calculated analytically.
Among these features are:
\begin{itemize}
\item The usage on a cluster has been facilitated. This has been
  achieved by decoupling the
  parts where Mathematica is needed completely from the numerical
  part, and by creating a directory structure such that the files for the numerical integration
  created by \secdec{} can be transferred to
  a cluster, where they are compiled and run. 
Run scripts are provided for the PBS, LSF and {\tt Condor} batch systems.
(The scripts for the LSF system are only available in the most recent version 3.0.9.)

The kinematics can be
  defined in a simple text file {\tt kinem.input} containing the numerical values for
  the invariants occurring in the integral, where the program will
  loop over all kinematic points defined in {\tt kinem.input}.
This new way to define the kinematic points to be calculated is 
also available in single machine mode.
\item Integrals with numerators can be calculated not only as
  contracted tensor integrals, but also by defining the numerators as
  inverse propagators.  Further, the list of propagator powers can
  also contain zero entries. 
 These features allow an easy interface to integral reduction programs
 as e.g. {\tt Reduze}~\cite{vonManteuffel:2012np}, where the propagator list defines the
 integral family, while the list of propagator powers defines the
 considered integral within this family. 
\item Two new decomposition strategies, based on algebraic geometry~\cite{Kaneko:2009qx,Kaneko:2010kj},
have been implemented. These strategies by construction cannot run into an infinite recursion 
at the iterated decomposition. Remarkably, the strategy {\tt G2} 
often leads to a lower number of subsectors than the ``default'' strategy {\tt X}.
This new feature is described in detail in~\cite{Schlenk:2016cwf}. 
\item The masses contained in the propagators can also be complex. 
For details we also refer to~\cite{Schlenk:2016cwf}. 
\item Integrals with propagators which are only linear in the loop
  momentum can be calculated as well. If these integrals are
  non-Euclidean, care has to be taken to ensure the correct sign for
  the contour deformation. The program will assume the same $i\delta$
  prescription as for Feynman propagators which are quadratic in the
  loop momenta.
\item A new treatment of spurious poles which have a worse than
  logarithmic divergence behaviour (related e.g.  to factors of
  $x^{-2-\eps}$ in the decomposed integrand) is available, which uses
  integration by parts (IBP) to remap these poles to logarithmic
  poles.
This option is invoked if the parameter {\tt IBPflag} is nonzero. 
\item The user can define thresholds below which the imaginary part
  of the integral is known to be zero; in this case \secdec{} will not
  attempt to calculate the imaginary part. This can speed up the
  calculation considerably.
\item For the numerical integration, the user can also choose to use
  an integrator from Mathematica. 
For one-dimensional integrals, the program automatically switches to
{\sc cquad}~\cite{cquad}, being faster and more accurate than Monte Carlo integration.
\item The user interface has been facilitated, decoupling completely 
the following three types of input:
(a) definition of the graph name, desired expansion order in $\eps$
and parameters for the numerical integration ({\tt param.input}), 
(b) graph definition ({\tt math.m}), 
(c) definition of the kinematics ({\tt kinem.input}).
\end{itemize}

The next release of \secdec{} will also contain the possibility to
create a library containing the sector functions produced by the program, 
such that the integrals can be linked to the calculation of a full
amplitude. This will allow to adjust the precision to which an individual
integral is calculated according to its importance within the full
amplitude.

Finally, for \secdec-4.0, the algebraic part will be completely
restructured, and will be implemented in \python{} rather than {\tt
  Mathematica}. This will be described in more detail in the
following subsection.

\subsection{Implementation in {\tt python}}

The forthcoming implementation of the algebraic part of \secdec{} in
\python{} has several advantages. One of them is certainly given by the
fact that \python{} is an open source software. 
In addition, the new implementation offers much more modularity. 
The various steps the program performs, like construction of the graph
polynomials, iterated decomposition, subtraction, $\eps$-expansion
(see also Fig.~\ref{fig:structure}), are available in the form of
\python{} modules and can be called separately. 
The decomposition can act on products of polynomials raised
to some power, so it is not limited to a loop integral structure 
of the form ${\cal F}^{\mathrm{expoF}}{\cal U}^{\mathrm{expoU}}$ times a numerator (see
Eq.~(\ref{eq:FU})).
Further, the new implementation allows to perform an expansion in
several regulators, not only the dimensional regulator $\eps$. 
The additional regulator could stem for example from analytic
regularisation~\cite{Becher:2011dz} or from parameter integrals containing
the string tension~\cite{Puhlfuerst:2015gta}.

An alpha-version of the python implementation can be found at 
{\tt http://secdec.hepforge.org}, where a manual is also linked. 





\section{Applications}

\subsection{Two-loop four-point integrals with two mass scales}
\label{sec:hh}

\begin{figure}[htb!]
\begin{center}
\includegraphics[width=0.4\textwidth]{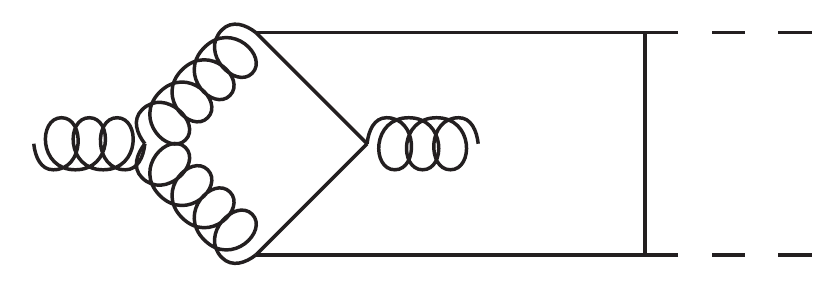}
\end{center}

\caption{Example of a non-planar two-loop box diagram with two mass scales. 
Solid lines denote massive propagators ($m_t$). 
The dashed lines denote massive legs with a mass ($m_H$) different from the internal mass.
	\label{fig:hhnp2b} }
\end{figure}

\begin{figure}[htb!]
\begin{center}
\includegraphics[width=0.75\textwidth,angle=0]{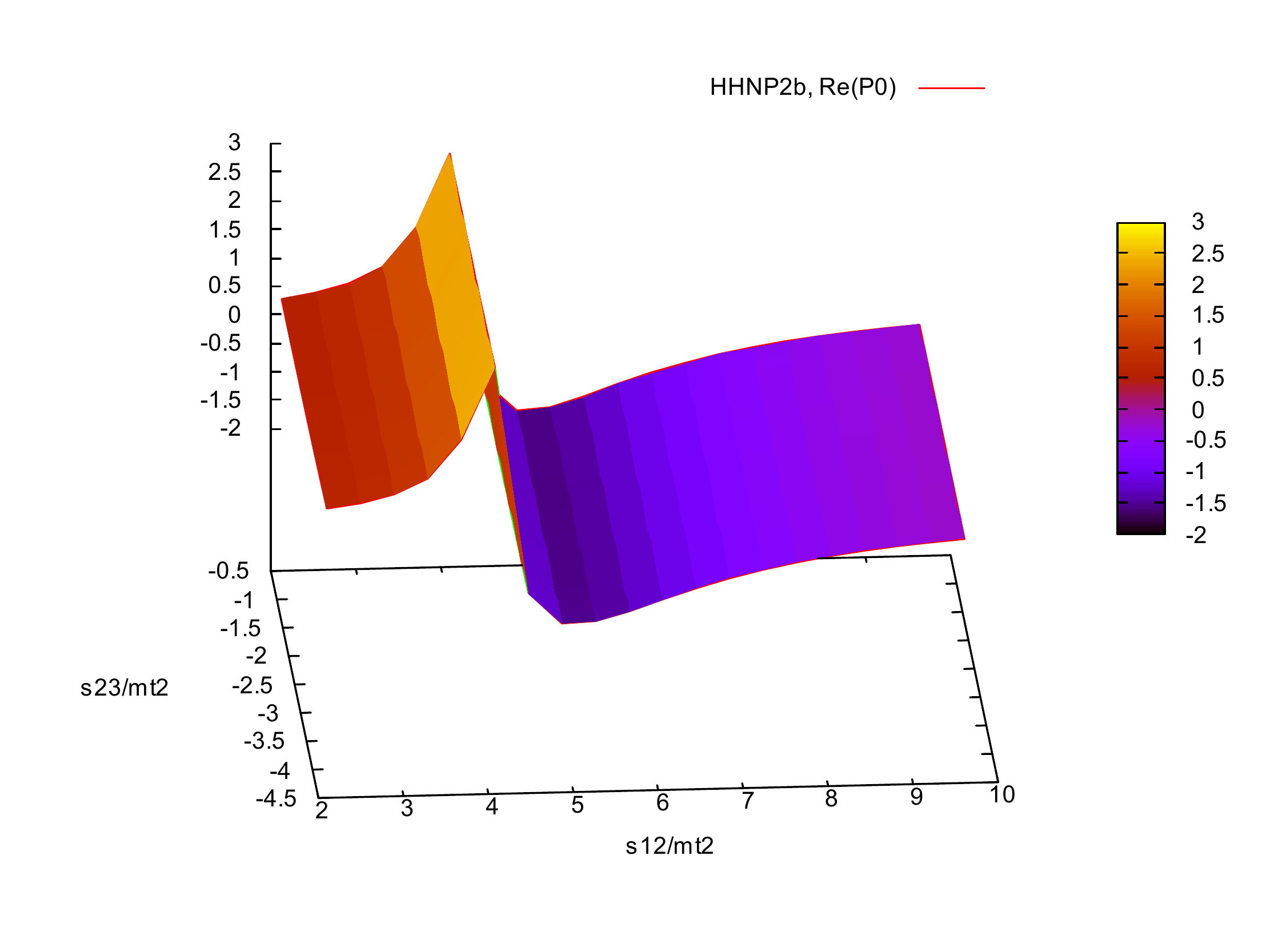}
\end{center}
\caption{Real part of the finite part of the graph shown in 
Fig.~\ref{fig:hhnp2b}, for the values $m_t=$173\,GeV, $m_H=$125\,GeV, $u=-s-t+2m_H^2$. \label{fig:resultshhnp2b} }
\end{figure}

As an example for an integral where the analytic result is unknown, 
we give results for a non-planar two-loop four-point function (see Fig.~\ref{fig:hhnp2b})  
containing both internal masses ($m_t$) and massive external legs with
a different mass ($m_H$).
The result for this diagram is of the form $I=P_1/\eps+P_0$.
Numerical results in the form of an interpolated scan over 20 values for $s_{12}/m_t^2$ times 20 values for $s_{23}/m_t^2$ 
for the real part of the finite term $P_0$ are shown in Fig.~\ref{fig:resultshhnp2b}.

\section{Conclusions}
We have presented new features of the program \secdec{}, 
which is a program to factorise poles from dimensionally regulated
parameter integrals, and to evaluate the pole coefficients
numerically. 
The embedding of \secdec{} into \gosam-2loop, a general framework for
the calculation of 2-loop amplitudes, has also been briefly described.
The new features of \secdec{} and further developments of these programs  aim at the usage of \secdec{}
within the calculation of full two-loop virtual amplitudes in cases where the analytic 
result for most of the master integrals is unknown, 
rather than the calculation of just a few kinematic points to check an analytic calculation of individual integrals.
This development is in line with the rapid developments in high performance computing.

\ack{
GH would like to thank the organizers of ACAT2016 for the nice conference. 
This research was supported in part by the 
Research Executive Agency (REA) of the European Union under the Grant Agreement
PITN-GA2012316704 (HiggsTools).
S. Borowka gratefully acknowledges financial support by the ERC
Advanced Grant MC@NNLO (340983).
}

\section*{References}

\providecommand{\newblock}{}

\end{document}